\documentclass[slac_one]{revtex4}
\usepackage{graphicx}
\usepackage{fancyhdr}
\usepackage{subfigure}
\usepackage{epsfig}
\usepackage[font=small,format=plain,labelfont=bf,up]{caption}
\begin{document}

\title{Minimum Bias Triggers at ATLAS, LHC} %% Paper title goes here

\author{Regina Kwee (on behalf of the ATLAS Collaboration)}
\affiliation{CERN, Switzerland/Humbolt-Universit\"at zu Berlin, Germany}

\begin{abstract}
In the first phase of LHC data-taking ATLAS will measure the charged-particle density at the initial center-of-mass energy of 10 TeV and then at 14 TeV. This will allow us improve our knowledge of soft QCD models and pin-down cross-sections of different classes of inelastic collisions at LHC energies. In particular, the dominant non-diffractive interaction is a key process to understanding QCD backgrounds when we reach higher luminosities. We highlight two minimum-bias triggers, sensitive to particles in complementary ranges in pseudo-rapidity, one based on signals from the Inner Detector, the other explicitly designed to trigger on inelastic processes. Studies of their trigger efficiencies as well as possible trigger biases are presented.
\end{abstract}

\maketitle

\thispagestyle{fancy}

\section{INTRODUCTION AND MOTIVATION FOR MINIMUM BIAS TRIGGERS} \label{common_intro} % Section title should be in all capitals. 
The cross-section predictions at LHC-energies vary around 50~\% due to theoretical uncertainties. The total cross-section is estimated to  $\sigma_{\text{tot}} \approx 100$ mb depending strongly on Monte Carlo models and different theoretical estimates~\cite{mc-ref}. In particular, at LHC-energies the particle multiplicities are unknown. Contributions to $\sigma_{\text{tot}}$ come from elastic and inelastic processes, where the inelastic part dominates with $\sigma_{\text{inel}} \approx 80$~mb. The leading inelastic processes are non-diffractive interactions with $\sigma_{\text{nd}} \approx 55$ mb, producing mainly low-p$_T$ particles in the central detector region. They are generally associated with minimum bias events, but technically the exact definition of a minimum bias event is detector dependent and made by the trigger system. Further inelastic contributions arise from single and double diffractive dissociations, with $\sigma_{\text{sd}} \approx 14$ mb and $\sigma_{\text{dd}} \approx 10$ mb respectively, producing particles mostly in the forward direction. Merely for completeness, so called central-diffractive dissociations are mentioned where a proton leaves a high diffractive mass also in the central detector part. This kind of interaction is not relevant for understanding the dominant QCD background events as their cross-section with $\sigma_{\text{cd}} \approx 1$ mb is expected to be negligibly small.
These inelastic soft interactions were often approximated in a phenomenological frame, their measurement will advance the theory for different soft QCD models. To improve our knowledge of the physics of these dominant processes, ATLAS~\cite{atlas-ref} will measure them at low luminosities, where events still contain single pp-interactions. This is essential for understanding the \emph{pile-up} of collisions they form at high luminosities.\newline 
ATLAS will use two Minimum Bias Triggers complementary in pseudo-rapidity to make these measurements. One is based on signals of the Inner Detector (ID), the other one is a dedicated experimental setup consisting of Minimum Bias Trigger Scintillators (MBTS). In these studies we investigated the initial scenario for data-taking. With a bunch pitch of 75 ns and a luminosity of L~$= 10^{31} \text{cm}^{-2}\text{s}^{-1}$, roughly 90~\% of the events are empty: the beam bunches penetrate each other without interacting, but leaving only electronic noise. They represent the main background to be rejected. After the commissioning phase, the minimum bias triggers also need to be efficient in suppressing beam-gas and halo events.

\section{MINIMUM BIAS TRIGGER SCINTILLATORS} \label{MBTS_intro}

The MBTS consist of 16 counters per side, of which 8 form an inner ring and another 8 an outer ring covering in total a pseudo-rapidity range of $2.1 < |\eta| < 3.8$ in the forward direction of each beam. They are mounted at $|z| = 3.6$~m between the ID and the Liquid-Argon-Calorimeter. The signals are read out through the electronics of the hadronic calorimeter~\cite{atlas-ref} which broaden the fast scintillator signals by roughly a factor of 50. A second trigger path has been installed to take advantage of the good time resolution of the scintillators which will allow to distinguish pp-collisions from beam-induced background events. The MBTS, as a relatively simple and robust detector system and sensitive to a minimum of detector activity, play a key role for detector commissioning~\cite{messina-ref}. After 3 to 4 months of running at luminosities of L~$ \approx 10^{33} \text{cm}^{-2}\text{s}^{-1}$, major radiation damage is expected.\newline
\begin{center}
\parbox[t]{0.36\textwidth}{
  \includegraphics[width=\linewidth]{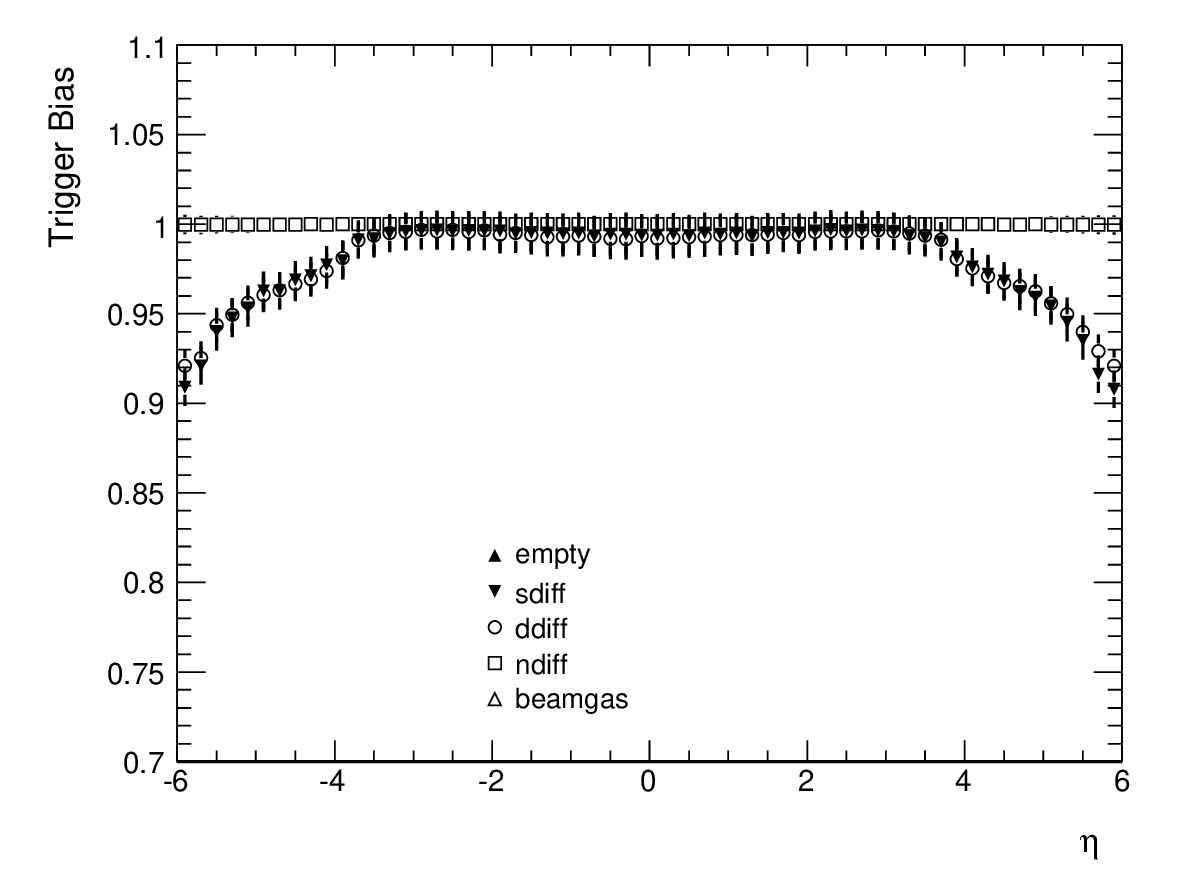}
  \captionof{figure}{Trigger bias for a requirement of any 2 hits in total.}
  \label{fig:biasMbts2}
}
\qquad
\parbox[t]{0.36\textwidth}{
  \includegraphics[width=\linewidth]{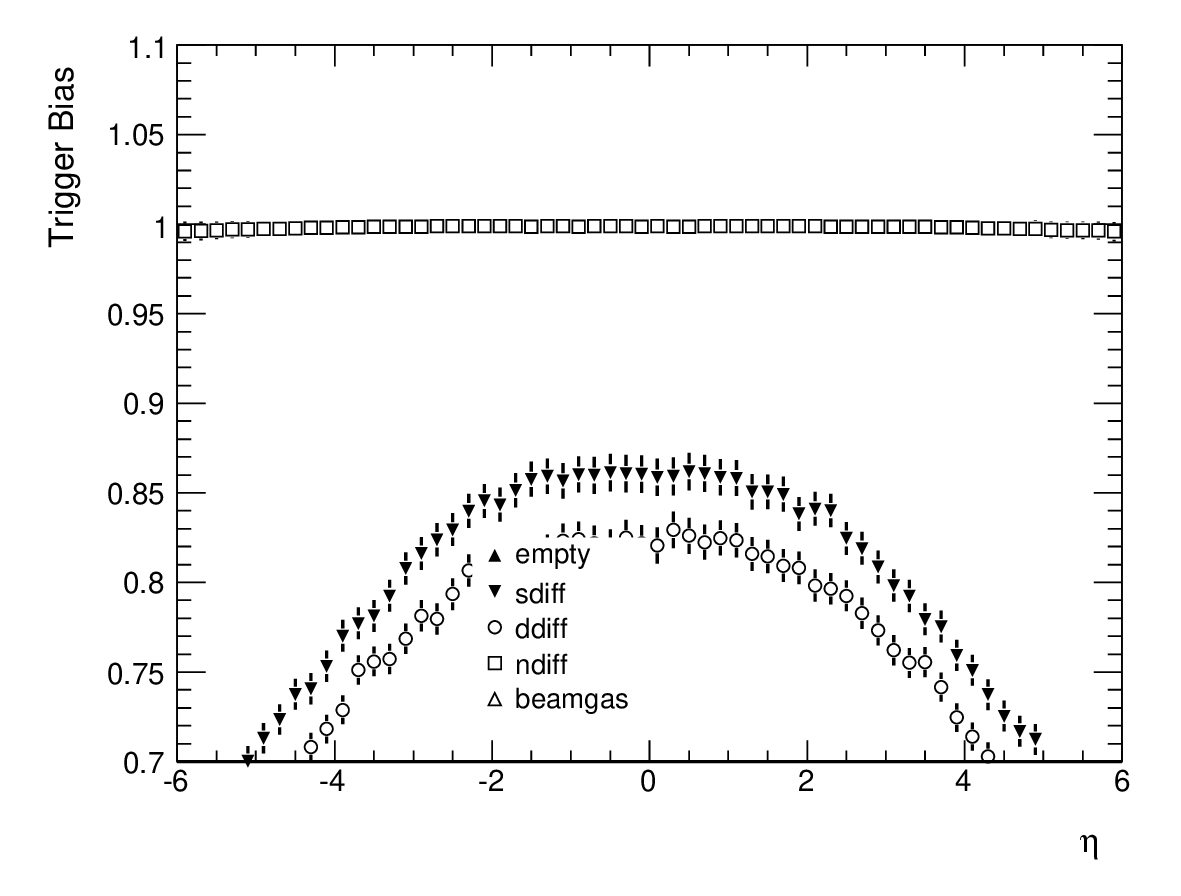}
  \captionof{figure}{Trigger bias for a requirement of a 1-hit coincidence.}
  \label{fig:biasMbts11}
}
\end{center}

\parindent0cm \textbf{Performance of Minimum Bias Trigger Scintillators}\\
To efficiently suppress empty events several trigger configurations have been studied for the MBTS requiring either a hit coincidence or total hit multiplicity. The efficiency was defined as the ratio of reconstructed events passing the trigger to the total number of reconstructed events and was computed for several simulated event samples. Due to their expected significance, non-diffractive events served as signal and empty events as background. Less significant but still not negligible, single and double diffractive samples were also classified as signal and beam-gas events were considered as background.
With a trigger threshold set to 40 mV, noise events can be highly suppressed while the non-diffractive events are still very efficiently retained with an expected signal amplitude of around 100 mV. A difference of these configurations is visible for forward directed events. A total hit multiplicity requirement would pass $\sim 30$~\% more of the diffractive events, but also 15~\% of the beam-gas events. The trigger bias was computed using the same event samples, but taking into account only truth quantities with a trigger threshold fixed to 40 mV for all counters. To estimate the bias, the ratio of stable, primary charged truth particles passing the trigger requirements and the same particles from all events was calculated. Comparing the two configurations from Fig.~\ref{fig:biasMbts2}, the total hit requirement seems much more favorable. The bias is flat for all event types over a large $\eta$-region up to the coverage of the MBTS of $|\eta| < 3.8$. Requiring a hit-coincidence as shown in Fig.~\ref{fig:biasMbts11} seems to bias both kinds of diffractive events strongly.

\section{INNER DETECTOR BASED MINIMUM BIAS TRIGGER}

Based on signals from the silicon sub-detectors of the ID, this minimum bias trigger covers a pseudo-rapidity range of $|\eta| < 2.5$. The innermost tracking devide in ATLAS is built of pixel sensors with about 80 million channels. It is surrounded by a silicon strip-detector, the Semi-Conductor-Tracker (SCT), with a read-out limited to roughly 6 million channels. The pixel hits provide already 3-dimensional position information forming \emph{SpacePoints}. In contrast, the SCT forms its SpacePoints by requiring a coincidence of two hits in strips that overlap in polar and azimuthal coordinates $\eta$ and $\phi$. The SCT provides thereby an intrinsic noise suppression. However, to keep flexibility, both pixel and SCT signals are exploited for the trigger. This trigger is configured to start off a random trigger at Level-1 (L1) to introduce no bias in the event selection. At the second trigger level (L2) detector activity is measured by counting the SpacePoints in the entire silicon-detector, which will be used to reject noise events. To discard beam-gas events, silicon tracks are reconstructed at the third trigger level, the Event Filter (EF), and tracks close to the interaction point are counted.

%% --- efficiency ID MB Trigger 
\begin{center}
  \parbox[t]{0.36\textwidth}{
    \includegraphics[width=\linewidth]{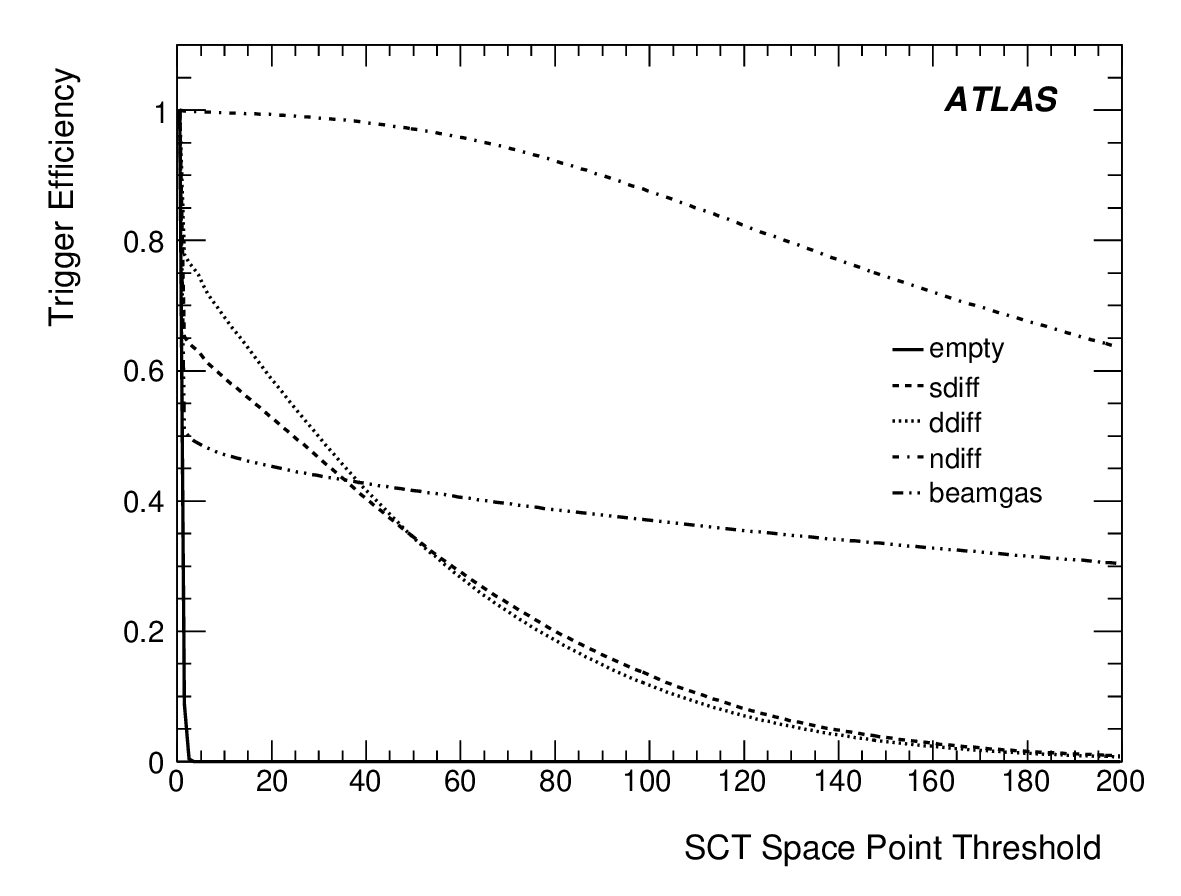}
    \captionof{figure}{Trigger efficiencies for a 3 SCT SpacePoints requirement. It drops quickly for empty events.}
    \label{fig:effSctSp}
  }
  \qquad
  \parbox[t]{0.36\textwidth}{
    \includegraphics[width=\linewidth]{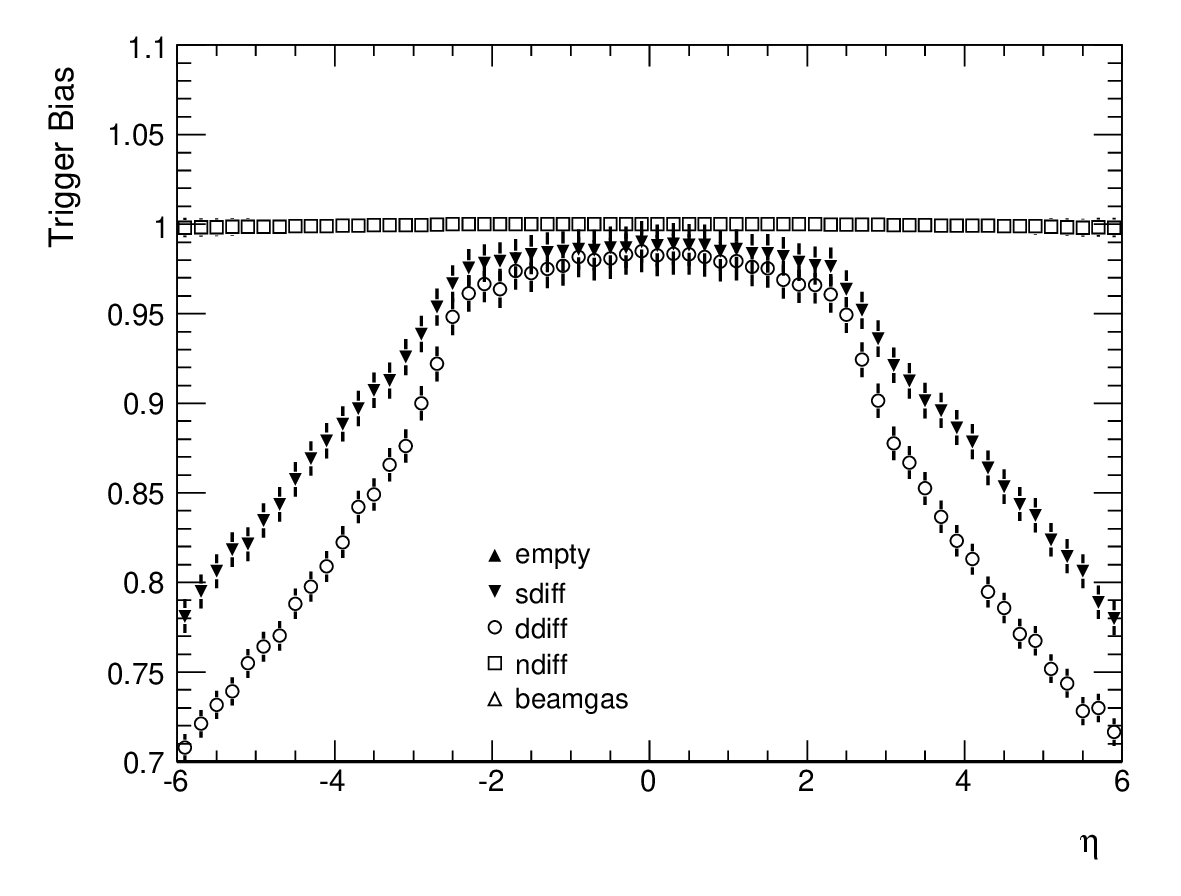}
    \captionof{figure}{Trigger bias for the requirement of 3 SCT and 12 pixel SpacePoints at L2 and 2 tracks with a $|z_0| < 200$~mm at EF.}
    \label{fig:biasSpTracks}
  }
\end{center}

\parindent0cm \textbf{Performance of the Inner Detector based Minimum Bias Trigger}\newline
The trigger performance was studied with the same event samples as for the MBTS. The trigger efficiency for requiring a minimum number of SCT SpacePoints at L2 is shown in Fig.~\ref{fig:effSctSp}. From the simulations we can expect that noise events can be strongly suppressed while keeping almost 100 \% of the non-diffractive events with a cut of e.g. 3 SCT SpacePoints. A similar plot has been made for the pixel SpacePoint requirement and showed comparable efficiencies for a cut at 12 pixel SpacePoints. The trigger efficiency was also computed for both SpacePoints requirements at L2 and an additional track requirement at the EF. The event is only accepted at the EF if a certain number of tracks could be reconstructed with a longitudinal impact parameter $|z_0| < 200$~mm. Setting the trigger threshold to 2 tracks would reject another 10 \% of the total beam-gas events, achieving a total rejection of more than 60~\%. In order to retain most of the diffractive events, this trigger condition should not be largely increased. However, for the non-diffractive sample even requiring 10 tracks leaves the trigger highly efficient with almost 100~\%.
The trigger bias was also calculated from the truth $\eta$-distributions having in the numerator the truth tracks which contain at least 12 pixel and 3 SCT SpacePoints and the total number of tracks in the denominator. Adding the track requirement of 2 tracks did not to influence the bias at all. The bias for both L2 and EF requirements is shown in Fig.~\ref{fig:biasSpTracks}. We see that the non-diffractive events are selected in an unbiased way, while a small bias can be observed in the diffractive samples towards the end of the ID acceptance ($\eta \sim 2.5$). Once, higher luminosities of L~$ \approx 10^{33} \text{cm}^{-2}\text{s}^{-1}$ are reached, a pure random trigger becomes more important.\newline

\textbf{Acknowledgements}\newline
The work presented was a collaboration of many people, especially those joining the Minimum Bias Trigger and Minimum Bias Working Group. The credits go also to W.H. Bell (University of Glasgow), D.Berge (CERN), C.Buttar (University of Glasgow), B.Demirkoz (CERN), E.Feng (University of Chicago), J.Husten (Michigan State University), N.Ellis (CERN), B.T.Martin (Michigan State University), A.Messina (CERN), S.Kama (DESY), M.Leyton (LBNL) and P.Steinberg (BNL). Particulary, I would like to thank William Bell, Ricardo J. Goncalo (Royal Holloway University of London), David Berge and Andrea Messina for their valuable input to the poster of this proceedings and this paper itself. Also, I would like to thank the organisers of ICHEP08 for the opportunity to present this work.


\begin{thebibliography}{9}   % Use for  1-9  references
%\begin{thebibliography}{99} % Use for 10-99 references
\footnotesize{
%% \bibitem{cscnote-ref}
%% Minimum Bias Working Group,``CSC Note of the Minimum Bias Working '', to be published, October 2008

\bibitem{mc-ref}
J.P. Guillaud and A. Sobol, ``Simulation of diffractive and non-diffractive processes at the LHC energy with PYTHIA and PHOJET MC event generators '', LAPP-EXP 2004-06, July 2004

\bibitem{atlas-ref}
ATLAS Collaboration, ``The ATLAS Experiment at the CERN Large Hadron Collider'', 2008 JINST 3 S08003, 14.Aug. 2008

\bibitem{messina-ref}
A. Messina et al., ``Commissioning of the ATLAS Level-1 Central Trigger system'', talk at TWEPP-08, 19.Sep. 2008

}
\end{thebibliography}
\end{document}